\documentclass[aps,preprint,showpacs,superscriptaddress,repreint]{revtex4-1}

\usepackage{graphicx}
\usepackage{dcolumn}
\usepackage{bm}
\usepackage{amsmath}
\usepackage{hyperref}
\usepackage{color}

\makeatother
\usepackage{tikz}
\begin{document}

\title{Effects of hydrodynamic interactions on rectified transport of self-propelled particles}

\author{Bao-quan  Ai} \email[Email: ]{aibq@scnu.edu.cn}
\affiliation{Guangdong Key Laboratory of Quantum Engineering and Quantum Materials, School of Physics and Telecommunication
Engineering, South China Normal University, Guangzhou 510006, China.}
\author{Ya-feng He}   \email[Email: ]{heyf@hbu.edu.cn}
\affiliation{College of Physics Science and Technology, Hebei University, Baoding 071002, China.}
 \author{Wei-rong Zhong} \email[Email: ]{wrzhong@jnu.edu.cn}
 \affiliation{Siyuan Laboratory, Guangzhou Key Laboratory of Vacuum Coating Technologies and
New Energy Materials, Department of Physics, Jinan University, Guangzhou 510632, China.}

\date{\today}

\begin{abstract}
 \indent Directed transport of self-propelled particles is numerically investigated in a three-dimensional asymmetric potential. Beside the steric repulsive forces, hydrodynamic interactions between particles have been taken into account in an approximate way. From numerical simulations, we find that hydrodynamic interactions can strongly affect the rectified transport of self-propelled particles. Hydrodynamic interactions enhance the performance of the rectified transport when particles can easily pass across the barrier of the potential, and reduce the rectified transport when particles are mainly trapped in the potential well.
\end{abstract}

\pacs{05. 40. -a, 82. 70. Dd}

\maketitle

\section{Introduction}
\indent The rectification of noise leading to unidirectional motion
in ratchet systems has been an active field of research over the last decade\cite{Hanggi}. Recently, a new class of active ratchet systems has
been realized through the use of active matter. Unlike passive Brownian ratchets, active ratchets do not require the application of an external driving force to produce rectification\cite{Marchetti0,Reichhardt0,Cates}. Experimental studies \cite{Galajda,Leonardo,Kaiser1,Koumakis0,Sanchez,Schwarz-Linek,Sokolov,Bricard,Mijalkov1,Kummel,Guidobaldi} have shown the key role of self-propulsion for rectifying particle motion in different structures, such as an array of asymmetric funnels \cite{Galajda}, a nano-sized ratchet-shaped wheel \cite{Leonardo},  the asymmetric barriers\cite{Koumakis0}, the microtubule bundles\cite{Sanchez}, the microscopic gears\cite{Sokolov}, and the nanoliter chambers\cite{Guidobaldi}. Active ratchet effects and variations on them will open a wealth of possibilities such as sorting, cargo transport, or micromachine construction.

\indent There has been increasing interest in theoretical work on rectification of self-propelled particles \cite{Wan,Ghosh,Ghosh1,angelani,angelani1,angelani2,potosky,Potiguar,Koumakis,McDermott,Lambert,Chen,Drocco,Li,Reichhardt1,Reichhardt2,Berdakin,Ai,Ai1,Mijalkov,Fily}.
The rectification phenomenon of overdamped swimming bacteria was theoretically observed in a system with an array of asymmetric barriers \cite{Wan}. In a compartmentalized channel, Ghosh et. al. \cite{Ghosh} performed simulations of active Janus particles in an asymmetric channel and found that the rectification can be orders of magnitude stronger than that for ordinary thermal potential ratchets. Angelani and co-workers \cite{angelani} studied the run-and-tumble particles in periodic potentials and found that the asymmetric potential produces a net drift speed. Potosky and co-workers \cite{potosky}
 found that the spatially modulated self-propelled velocity can induce the directed transport. McDermott et. al. found the collective ratchets and current reversals of active particles in quasi-one-dimensional asymmetric substrates\cite{McDermott}. The collection of bacteria is able to migrate against the funnel-shaped barriers by creating and maintaining a chemoattractant gradient \cite{Lambert}. Li and coworkers\cite{Li} manipulated transport of overdampd pointlike Janus particles in narrow two-dimensional corrugated channels. In addition, the chiral active particles can be rectified and sorted in complex and crowded environments \cite{Ai1,Mijalkov}.

\indent Most of the theoretical studies on ratchet transport of active particles neglect the long range hydrodynamic interactions. However, hydrodynamic interactions in ratchet systems could exhibit peculiar behaviors. Some theoretical works studied the effects of hydrodynamic interactions on ratchet transport of passive Brownian particles\cite{Grimm,Golshaei,Fornes}. Grimm and coworkers found that hydrodynamic interactions can significantly enhance the performance of passive Brownian ratchet when the ratchet states are changed individually\cite{Grimm}. Golshaei and Najafi studied the role of hydrodynamic interactions with walls on the rectified transport of passive particles and found that the long range hydrodynamic interactions with walls reduce the efficiency of the Brownian ratchet\cite{Golshaei}. However, the effects of hydrodynamic interactions on the ratchet performance of active Brownian particles (e. g. self-propelled particles) are not yet clear.

\indent In this paper, we studied the rectified transport of interacting self-propelled particles in a three-dimensional asymmetric substrate. Hydrodynamic interactions between particles are included in the Rotne-Prager-Yamakawa approximation. We focus on finding how hydrodynamic interactions influence the performance of the ratchet systems. It is found that hydrodynamic interactions enhance the performance of the active ratchet when the self-propelled force dominates the transport, and reduce the rectified transport when particles are mainly trapped in the potential well.

\section{Model and methods}
We consider $N$ interacting self-propelled particles moving in a three-dimensional box of size $L_x\times L_y\times L_z$ with periodic boundary conditions.
Each particle is represented by a hard sphere of the radius $a$. The dynamics of particle $i$ is characterized by the position $\mathbf{r}_i\equiv (x_i,y_i,z_i)$ of its center and the orientation vector $\mathbf{n}_i$. Hydrodynamic interactions between particles are considered by using the Rotne-Prager-Yamakawa tensor\cite{Yamakawa}
\begin{equation}\label{D1}
  \bm{\mu}_{ij}=\mu_0\mathbf{I}, \quad \text{if} \quad  i=j,
\end{equation}
\begin{equation}\label{D2}
  \bm{\mu}_{ij}=\mu_0\frac{3a}{4r_{ij}}\bigg[\bigg(1+\frac{2a^2}{3r_{ij}^2}\bigg)\mathbf{I}+\bigg(1-\frac{2a^2}{r_{ij}^2}\bigg)\frac{\mathbf{r}_{ij}\otimes\mathbf{r}_{ij} }{r_{ij}^2}     \bigg],\\\quad\text{if}\quad r_{ij}\geq 2a, i\neq j,
\end{equation}
\begin{equation}\label{D3}
  \bm{\mu}_{ij}=\mu_0\bigg[\bigg(1-\frac{9}{32}\frac{r_{ij}}{a}\bigg)\mathbf{I}+\frac{3}{32}\frac{r_{ij}}{a}\frac{\mathbf{r}_{ij}\otimes\mathbf{r}_{ij} }{r_{ij}^2}  \bigg],\\\quad\text{if}\quad r_{ij}< 2a, i\neq j,
\end{equation}
where $\mu_0=1/(6\pi\eta a)$, $\mathbf{r}_{ij}=\mathbf{r}_j-\mathbf{r}_i$, $r_{ij}=|\mathbf{r}_{ij}|$. $\eta$ is the solvent viscosity. $\mathbf{I}$ is the unity tensor and $\mathbf{r}\otimes\mathbf{r}$ denotes the dyadic product.

\indent The dynamics of particle $i$ with hydrodynamic interactions is described by the following Langevin equations \cite{Ermak}
\begin{equation}\label{e1}
  m_i\frac{d^2 \mathbf{r}_i}{dt^2} =-\sum_{j}\xi_{ij}\frac{d \mathbf{r}_i}{dt}+f_{0}{\mathbf{n}}_{i}+\mathbf{F}_{i}^r+\mathbf{F}_{i}^{sub}+\sum_{j}\sigma_{ij}\cdot \mathbf{W}_{j},
\end{equation}
\begin{equation}\label{e2}
  \frac{d{\mathbf{n}}_{i}}{dt}={\mathbf{n}}_{i}\times \bm{\eta}_{i},
\end{equation}
where $m_i$ is the mass of particle $i$. $f_0$ is the self-propulsion force.  $\bm{\eta}_{i}$ is a zero-mean Gaussian white-noise random vector with variances $\langle {\eta}_{\alpha,i}(t)\eta_{\beta,j}(t^{'})\rangle=2D_r\delta_{\alpha\beta}\delta_{ij}\delta(t-t^{'})$ with $\alpha,\beta=x,y,z$. $D_r$ is the rotational diffusion coefficient.  $\mathbf{W}_{j}$ is described by a Gaussian distribution with the mean and covariance $\langle \mathbf{W}_{i}\rangle=0$ and $\langle \mathbf{W}_{i}(t) \mathbf{W}_{j}(t^{'})\rangle=2\delta_{ij}\delta(t-t^{'})$.  The coefficient $ \sigma_{ij}$ is related to the hydrodynamic friction tensor $\xi_{ij}$ by $\xi_{ij}=\frac{1}{k_B T}\sum_{l}\sigma_{il}\sigma_{jl}$, where $k_B T$ is the thermal energy. We introduce the mobility tensor $\bm{\mu}_{ij}$ shown in Eqs.(\ref{D1},\ref{D2},\ref{D3}), which is related to $\xi_{ij}$ by $\sum_j \xi_{ij}\bm{\mu}_{jl}=\sum_j \bm{\mu}_{ij}\xi_{jl}=\delta_{il}$. $\delta $ is the delta function and $\langle...\rangle $ denotes an ensemble average over time.

\indent Particles are assumed to move in the ovderdamped low-Reynolds-number regime. Ignoring the inertia of the particle, Eq.(\ref{e1}) can be written as a first order stochastic differential equation with the help of $\bm{\mu}_{ij}$,
\begin{equation}\label{e5}
  d\mathbf{r}_i=\bigg [k_B T\sum_{j=1}^{N}\frac{\partial \bm{\mu}_{ij} }{\partial \mathbf{r}_j}+\sum_{j=1}^{N}{\bm{\mu}_{ij}\cdot(f_{0}{\mathbf{n}}_{j}+\mathbf{F}_{j}^r+\mathbf{F}_{j}^{sub})}\bigg ]dt+R_{i}(dt),
\end{equation}
where the displacement $R_i(dt)$ is a random displacement with a Gaussian distribution with the zero mean and covariance $\langle R_i(dt)R_j(dt)\rangle=2 k_B T\bm{\mu}_{ij} dt$. In general, these tensors are functions of the complete spatial configuration of all particles.

\indent For the Rotne-Prager-Yamakawa tensor, $\sum_{j}\ \frac{\partial \bm{\mu}_{ij} }{\partial \mathbf{r}_j}$ is always equal to zero. Thus, Eq. (\ref{e5}) is reduced to
\begin{equation}\label{e6}
  d\mathbf{r}_i=\bigg [\sum_{j=1}^{N}{\bm{\mu}_{ij}\cdot(f_{0}{\mathbf{n}}_{j}+\mathbf{F}_{j}^r+\mathbf{F}_{j}^{sub})}\bigg ]dt+R_{i}(dt).
\end{equation}

\indent $\mathbf{F}_{i}^r=\sum_{j\neq i} \mathbf{F}_{ij}$ is the total steric repulsive force on particle $i$. The interactions $\mathbf{F}_{ij}$ between the spherical particles of the radius $a$ are taken as short-ranged harmonic repulsive forces: $\mathbf{F}_{ij}=k(2a-r_{ij})\hat{\mathbf{r}}_{ij}$ if particles overlap ($r_{ij}<2a$), and $\mathbf{F}_{ij}=0$ otherwise. Here $k$ denotes the spring constant and $\hat{\mathbf{r}}_{ij}=\mathbf{r}_{ij}/r_{ij}$.  In order to mimic hard particles, we use large values of $k$, thus ensuring that particle overlaps decay quickly. The interactions between particles are radially symmetric and do not directly coupled to the angular dynamics.

\indent The substrate force $\mathbf{F}_{i}^{sub}=-\nabla V$ ($\nabla$ is the gradient operator) arises from the following periodic potential
\begin{equation}\label{hx}
    V(x,y,z)=U_0[\sin(\frac{2\pi }{l_x}x)+\frac{\Delta}{4}\sin(\frac{4\pi}{l_x}x)+\sin(\frac{2\pi}{l_y}y)+\sin(\frac{2\pi}{l_z}z)],
\end{equation}
where $l_{x,y,z}$ is the substrate period and $U_0$ is height of the potential. The potential is asymmetric in $x$ direction and symmetric in $y$ and $z$ directions. $\Delta$ is the asymmetric parameter of the potential in $x$ direction and the potential is completely symmetric at $\Delta=0$.

\indent By introducing the characteristic length scale
and the time scale: $\hat{\mathbf{r}}=\frac{\mathbf{r}}{a}$, $\hat{t}=\frac{t}{\tau}$ with $\tau$$=$$\frac{a^2}{k_B T \mu_0}$, Eq. (\ref{e6}) can be rewritten in the following forms
\begin{equation}\label{f5}
  d\hat{\mathbf{r}}_i=\bigg [\sum_{j=1}^{N}\hat{\bm{\mu}}_{ij}\cdot(\hat{f}_{0}{\mathbf{n}}_{j}+\hat{\mathbf{F}}_{j}^r+\hat{\mathbf{F}}_{j}^{sub})\bigg ]d\hat{t}+\hat{R}_{i}(d\hat{t}),
\end{equation}
 and Eq. (\ref{e2}) can be rewritten as \cite{Winkler}
\begin{equation}\label{f6}
  \frac{d{\mathbf{n}}_{i}}{d\hat{t}}={\mathbf{n}}_{i}\times \hat{\bm{{\eta}}}_{i},
\end{equation}
where $\hat{\bm{\mu}}_{ij}=\frac{\bm{\mu}_{ij}}{\mu_0}$, $\hat{f}_0=\frac{f_0 a}{k_B T}$, $\hat{k}=\frac{ka^2}{k_B T}$, $\hat{U}_0=\frac{U_0}{k_B T}$, $\hat{D}_{r}=\frac{D_{r}a^2}{k_B T \mu_0}$, and $\hat{\chi}=\chi/a$ ($\chi=L_{x,y,z},l_{x,y,z}$). $\hat{\bm{{\eta}}}_i$ is a zero-mean Gaussian white-noise random vector with $\hat{D}_{r}$ variance. From now on, we will use only the dimensionless variables and omit the hat for all quantities appearing in the above equations.

\indent Because the potential in $y$ and $z$ directions is completely symmetric, the ratchet transport only occurs in the $x$ direction. To quantify the ratchet effect, we measure the average velocity in $x$ direction. The average velocity in the asymptotic long-time regime can be obtained from the formula
\begin{equation}\label{es}
v_x=\frac{1}{N}\sum_{i=1}^{N}\lim_{t\rightarrow\infty}\frac{\langle x_i(t)-x_i(0)\rangle}{t},
\end{equation}
  and we define the scaled average velocity $V_s=\frac{v_x}{\mu_0f_0}$ for convenience.
¡¡
\section{Results and Discussion}
\indent Unless otherwise noted, our simulations are under the parameter sets: $L_x=L_y=L_z=16\pi$, $l_x=l_y=l_z=2\pi$, and $k=100.0$.  We vary $N$, $U_0$, $D_{r}$, $\Delta$ and $f_0$ and measure the average velocity of self-propelled particles in $x$ direction. The results are shown in Figs. (1)-(4). 
The blue (red) lines show the simulation results when hydrodynamic interactions are neglected (included). 
\begin{figure}[htpb]
\vspace{1cm}
  \label{fig:asym_contour}
  \includegraphics[width=0.45\columnwidth]{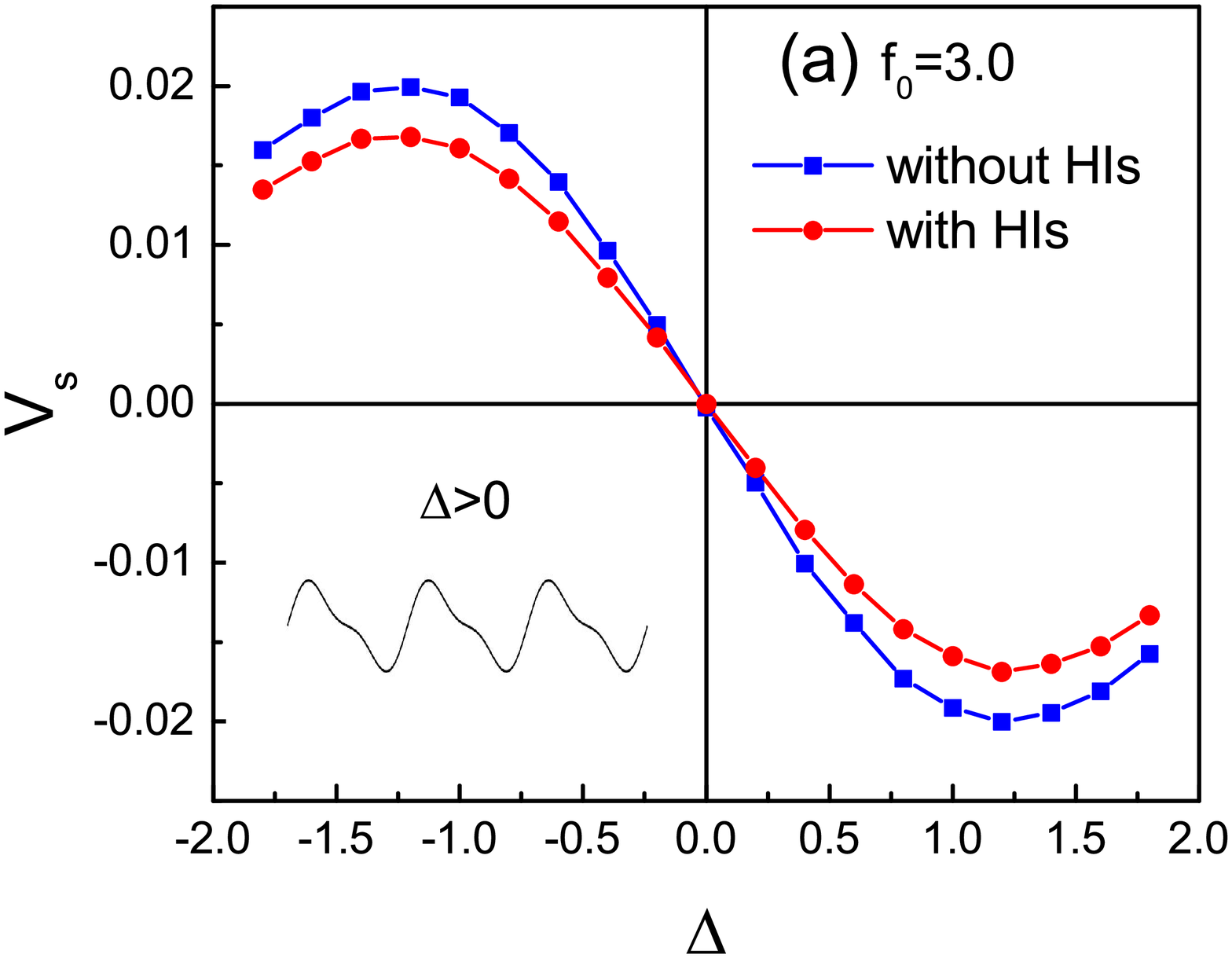}
  \includegraphics[width=0.45\columnwidth]{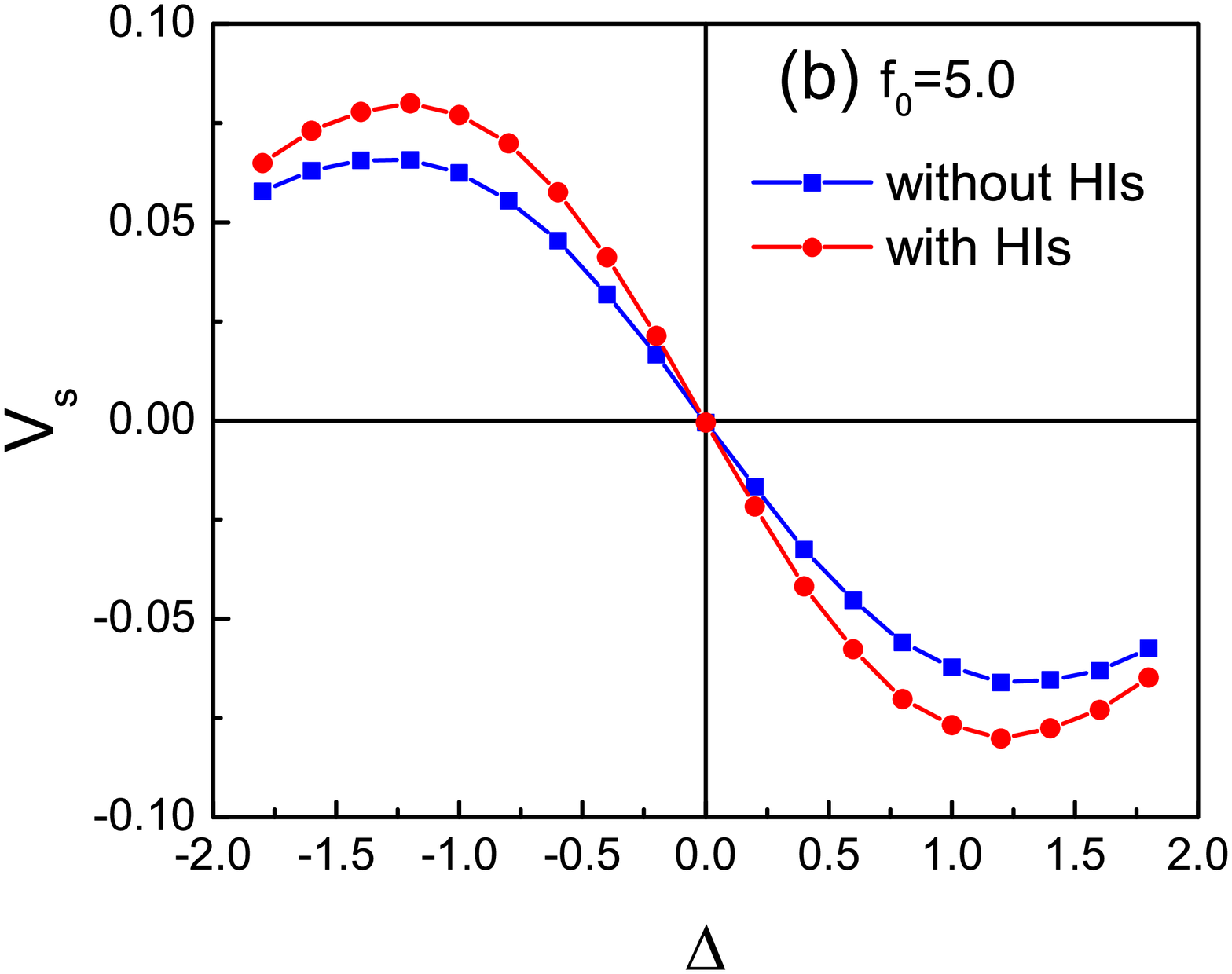}
  \includegraphics[width=0.45\columnwidth]{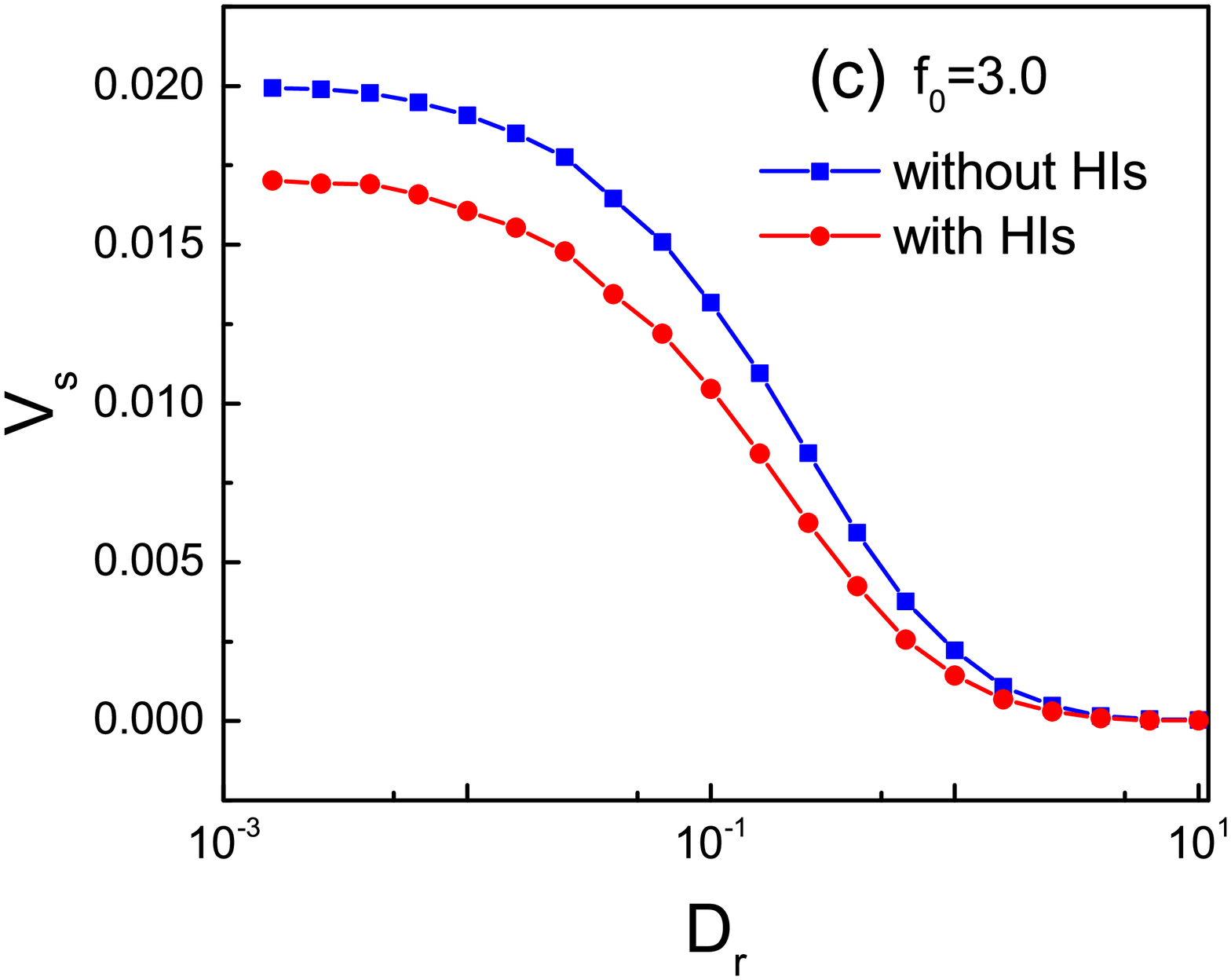}
  \includegraphics[width=0.45\columnwidth]{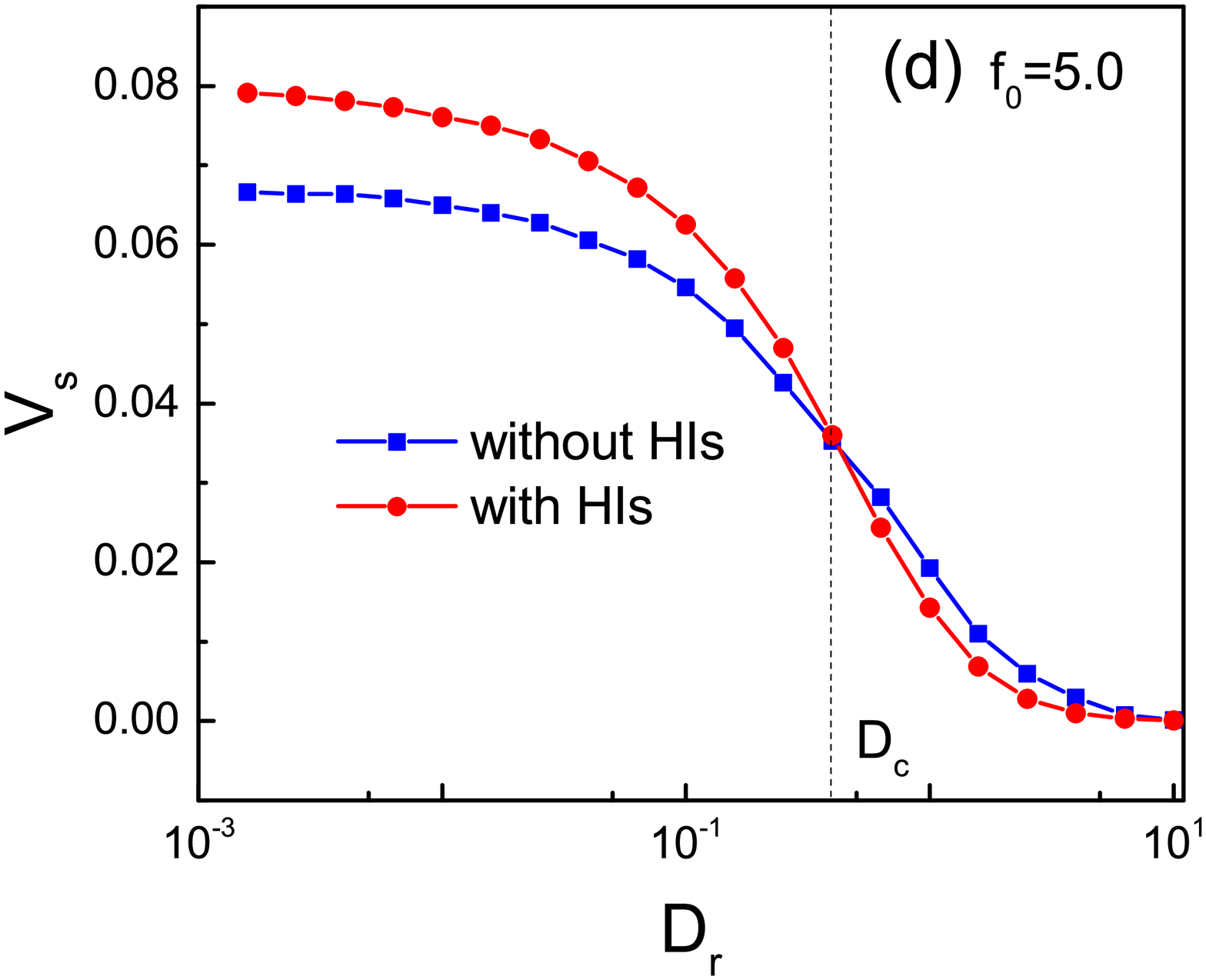}
  \caption{(Color online) Scaled average velocity $V_s$ when hydrodynamic interactions are included (with HIs) or neglected (without HIs) . (a) $V_s$  as a function of the asymmetric parameter $\Delta$ at $f_0=3.0$. (b)$V_s$ as a function of the asymmetric parameter $\Delta$ at $f_0=5.0$. (c)$V_s$ as a function of the rotational diffusion coefficient $D_r$ at $f_0=3.0$. (d)$V_s$ as a function of the rotational diffusion coefficient $D_r$ at $f_0=5.0$. The inset of Fig. 1(a) describes the profile of $x$ direction potential for $\Delta>0$. The other parameter are $N=10$, $\Delta=-1.5$, $D_r=0.01$, and $U_0=5.0$.
  \label{fig:Parallel_pressure_current}}
\end{figure}

\indent Figures 1(a) and 1(b) show the scaled average velocity $V_s$ as a function of the asymmetrical parameter $\Delta$ when hydrodynamic interactions are included as well as neglected. It is found that $V_s$ is positive for $\Delta<0$, zero at $\Delta=0$, and negative for $\Delta>0$. A qualitative explanation of this behavior can be given by the following argument. For $\Delta=0$ (symmetric case), the probabilities of crossing right and left barriers are the same, so there is a null net particles flow. For $\Delta>0$, the right side from the minima of the potential is steeper (shown in the inset of Fig. 1(a)), it is easier for particles to move toward the gentler slope side than toward the steeper side, so the average velocity is negative. In the same way, the average velocity is positive for $\Delta<0$. Therefore, the resulting direction of particles is completely determined by the asymmetric parameter $\Delta$ of the potential.

\indent Figures 1(c) and 1(d) show the scaled average velocity $V_s$ as a function of the rotational diffusion coefficient $D_r$. We find that the average velocity decreases monotonously with the increase of $D_r$ when hydrodynamic interactions are included as well as neglected.  When $D_r \rightarrow 0$, the self-propelled angle almost does not change, and the average velocity approaches its maximal value, which is similar to the adiabatic case in the forced thermal ratchet\cite{Magnasco}. When $D_r \rightarrow\infty$, the self-propelled angle changes very fast, the self-propulsion acts as a zero mean white noise and the system is in equilibrium, so the average velocity tends to zero.

\indent From Figs. 1(a)-1(d), we can conclude that the ratcheting behaviors are similar when hydrodynamic interactions are included as well as neglected. However, hydrodynamic interactions can strongly affect the performance of the rectified transport. These results can be classified into three cases: (1)$f_0<U_0$ for all values of $D_r$, the scaled average velocity without hydrodynamic interactions is larger than that with hydrodynamic interactions (shown in Figs. 1(a) and 1(c)), where the substrate force is dominated and particles are mainly trapped in the potential well. (2)$f_0\geq U_0$ and $D_r<D_c$, the scaled average velocity without hydrodynamic interactions is less than that with hydrodynamic interactions (shown in Figs. 1(b) and 1(d)), where particles can easily pass across the barrier of the potential and the self-propulsion force dominates the transport. (3)$f_0\geq U_0$ and $D_r>D_c$, the scaled average velocity without hydrodynamic interactions is larger than that with hydrodynamic interactions (shown in Fig. 1(d)), where the self-propelled angle changes very fast, the self-propulsion acts as a zero mean white noise and particles are mainly trapped in the potential well. Therefore, we can conclude that hydrodynamic interactions enhance the performance of ratchet transport when the self-propulsion force dominates the transport (particles can easily pass across the barrier of the potential), and reduce the rectified transport when the substrate force dominates the transport(particles are mainly trapped in the potential well).

\indent The following consideration gives a qualitative understanding of these behaviors. In general, hydrodynamic interactions between particles can induce synchronization.  When the substrate force is dominated ($f_0=3.0$), a few particles (named A) can pass across the barrier of the potential and most particles (named B) stay in the local minima of the potential. Due to the synchronization from hydrodynamic interactions, B pulls A to the local minima of the potential, more particles cannot pass across the barrier, thus the average velocity decreases. Similarly, when the self-propulsion is dominated ($f_0=5.0$), a few particles (named A) stay in the local minima of the potential and most particles (named B) can pass across the barrier. Because of the synchronization, B pushes A to leave the local minima of the potential, more particles can pass across the barrier, thus resulting in the increase of the average velocity. In order to confirm the above conclusion, we also investigate the dependence of the scaled average velocity $V_s$ on the parameters $U_0$ , $f_0$ and $N$ in Figs. (2-3).

\begin{figure}[htpb]
\vspace{1cm}
  \label{fig:asym_contour}\includegraphics[width=0.47\columnwidth]{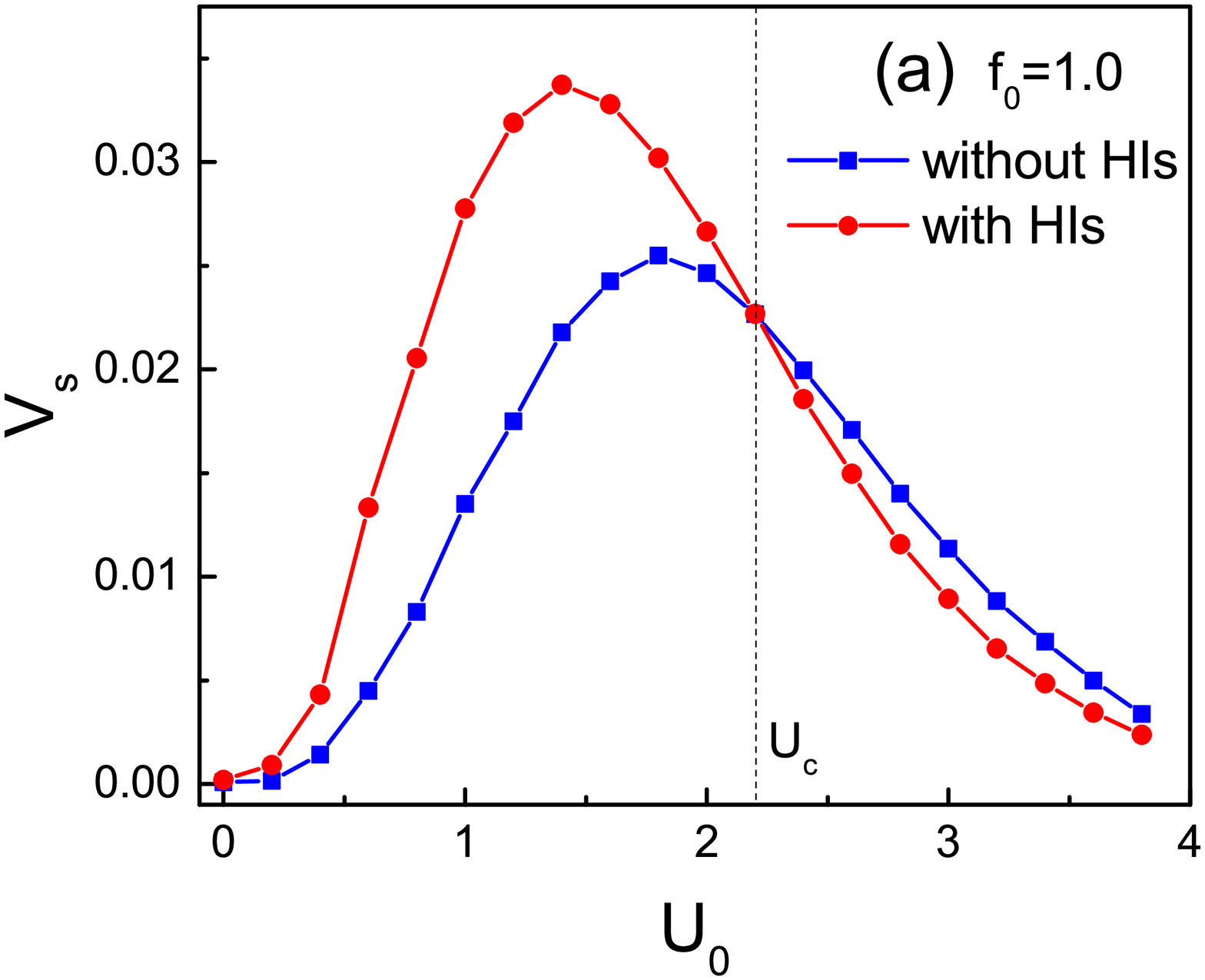}
  \includegraphics[width=0.47\columnwidth]{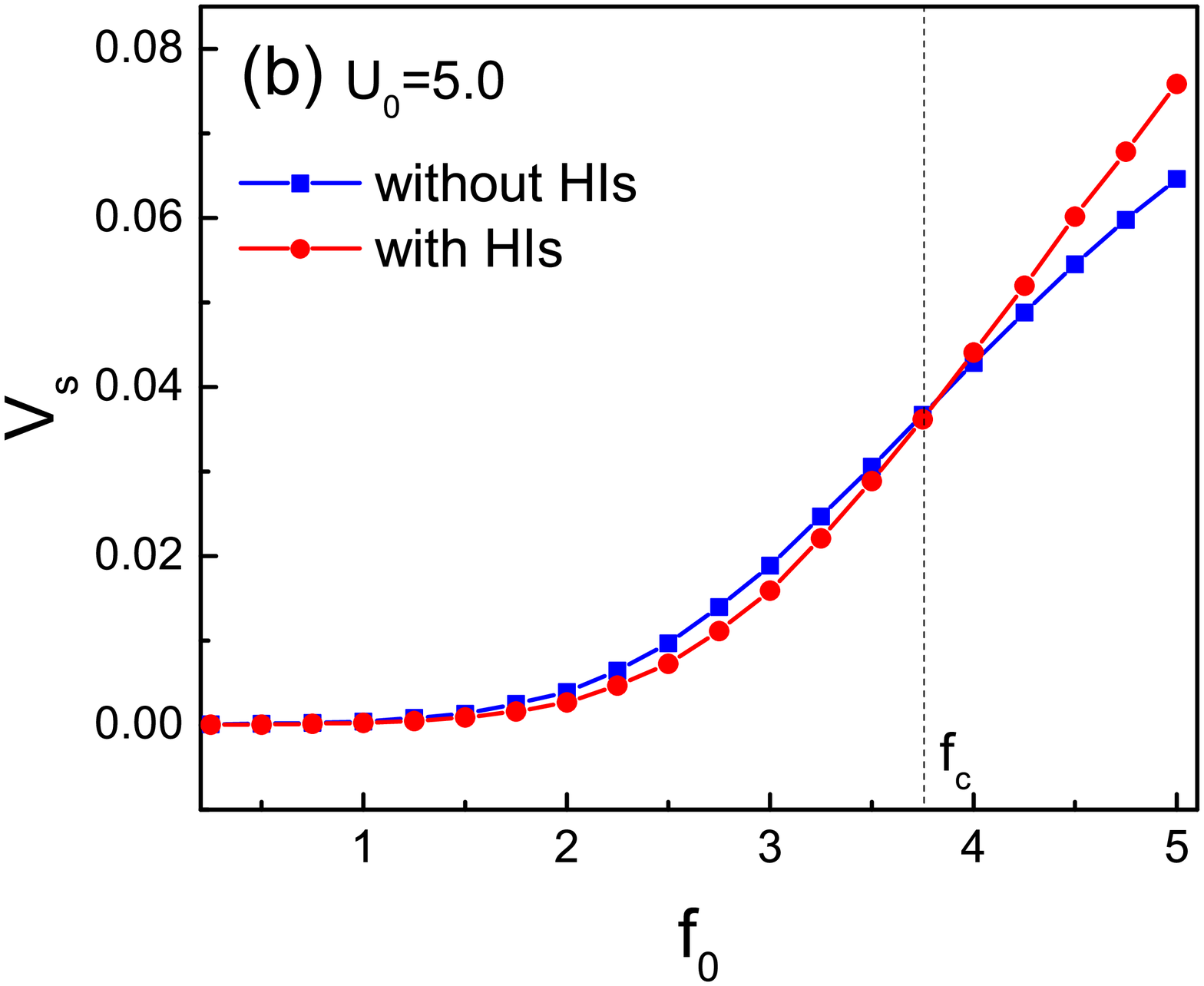}
  \caption{(Color online)(a)Scaled average velocity $V_s$ as a function of the potential height $U_0$ when hydrodynamic interactions are included (with HIs) or neglected (without HIs).
  (b)Scaled average velocity $V_s$ as a function of the self-propulsion force $f_0$ when hydrodynamic interactions are included (with HIs) or neglected (without HIs). The other parameters are $N=10$, $D_r=0.01$, and $\Delta=-1.5$.
    \label{fig:Parallel_pressure_current}}
\end{figure}

\indent
Figure 2(a) gives the scaled average velocity $V_s$ as a function of the potential height $U_0$ when hydrodynamic interactions are included as well as neglected. The curves are observed to be bell shaped. When $U_0\rightarrow 0$, the effects of the potential disappear and the scaled average velocity tends to zero. When $U_0\rightarrow \infty$, the particles cannot pass across the barrier and the average velocity also goes to zero. There exists an optimal value of $U_0$ at which the scaled average velocity $V_s$ is maximal. Compared with no hydrodynamic interactions, the position of the peak with hydrodynamic interactions shifts to the left. Similar to Fig. 1, hydrodynamic interactions improve the rectified transport when particles can easily pass across the barrier of the potential ($U_0<U_c$) and hydrodynamic interactions reduce the rectified transport when particles are mainly trapped in the potential well ($U_0>U_c$). The same conclusion can also be obtained from Fig. 2(b), where hydrodynamic interactions improve (reduce) the rectified transport when $f_0>f_c$ ($f_0<f_c$). Critical values of $U_c$ and $f_c$ are determined by the other parameters of the system.

\begin{figure}[htpb]
\vspace{1cm}
  \label{fig:asym_contour}\includegraphics[width=0.6\columnwidth]{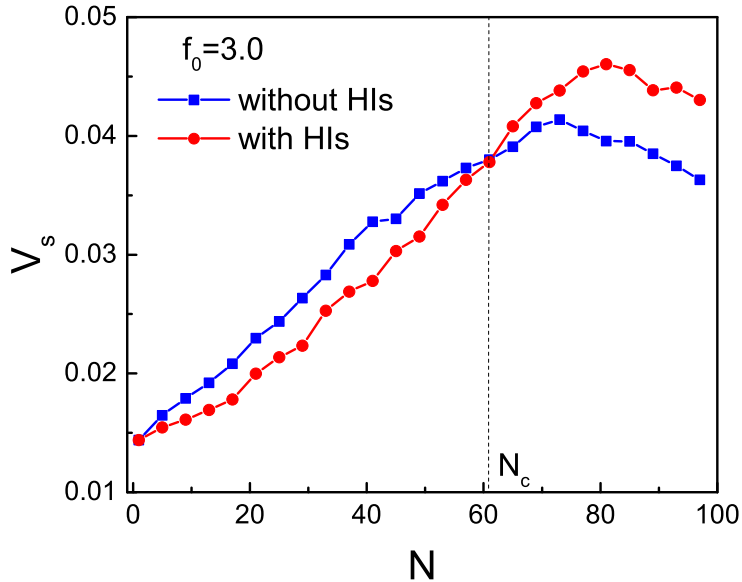}
  \caption{(Color online)Scaled average velocity $V_s$ as a function of particle number $N$ for both with HIs and without HIs at $f_0=3.0$, $U_0=5.0$, $D_r=0.01$, and $\Delta=-1.5$.
    \label{fig:Parallel_pressure_current}}
\end{figure}
\indent Figure 3 shows the scaled average velocity $V_s$ as a function of particle number $N$ when hydrodynamic interactions are included as well as neglected. The scaled average velocity $V_s$ is a peaked function of the particle number $N$.  The steric repulsive interactions between particles can cause two factors (A) reducing the self-propelled driving, which blocks the ratchet transport and (B)activating motion in an analogy with the thermal noise activated motion for a single stochastically driven ratchet, which facilitates the ratchet transport. For small $N$, the factor B first dominates the transport, so the average velocity
increases with $N$. However, for large $N$, the factor A dominates the transport, so the average velocity decreases with increasing $N$.  Therefore, there exists an optimal value of $N$ at which the scaled average velocity is maximal.

\indent From Fig. 3, we can also find the effects of hydrodynamic interactions on the performance of the ratcheting transport.  hydrodynamic interactions reduces the rectified transport for $N<N_c$ and improves the rectified transport for $N>N_c$. The critical $N_c$ is determined by the other parameters of the system. This can be explained as follows. In Fig. 3, we consider the case of $f_0=3.0$, where particles are mainly trapped in the potential well. For small value of $N$, a few particles can pass across the barrier, due to the synchronization, these particles are pulled to the local minima of the potential, thus the scaled average velocity reduces. For large value of $N$, the steric repulsive interactions between particles push particles to leave the local minima of the potential, most particles can pass across the barrier, thus the synchronization from hydrodynamic interactions can improve the rectified transport in this case.

\section {Concluding remarks}
\indent In this work, we numerically studied the rectified transport of self-propelled particles in a three-dimensional asymmetric potential. The long range hydrodynamic interactions between particles have been considered in the Rotne-Prager-Yamakawa approximation. The out-of-equilibrium condition of the active particle is an intrinsic property, which can break thermodynamical equilibrium and induce the directed transport.  The direction of transport is entirely determined by the asymmetric parameter of the potential, the scaled average velocity $V_s$ is positive for $\Delta<0$, zero at $\Delta=0.0$, and negative for $\Delta>0$. Although the ratchet behaviors are essentially similar when hydrodynamic interactions are included as well as neglected, hydrodynamic interactions can strongly affect the performance of the active ratchet systems. Hydrodynamic interactions enhance the performance of the active Brownian ratchet when particles can easily pass across the barrier of the potential, and reduce the rectified transport when particles are mainly trapped in the potential well. In addition, there exist optimal system parameters (particle number $N$ and the potential height $U$) at which the scaled average velocity takes its maximal value. We expect that our results can provide insight into out-of-equilibrium phenomena and have potential applications in artificial swimmers, nanobots, and other self-driven systems.

\section*{Acknowledgements}
\indent This work was supported in part by the National Natural Science Foundation of China (Grant Nos. 11575064 and 11175067), the PCSIRT (Grant No. IRT1243), the GDUPS (2016), and the Natural Science Foundation of Guangdong Province (Grant No. 2014A030313426).

\end{document}